# Signal Processing with Pulse Trains: An Algebraic Approach- Part II

Gabriel Nallathambi and Jose C. Principe

*Abstract*—The integrate and fire converter (IFC) enables an alternative to digital signal processing. IFC converts analog signal voltages into time between pulses and it is possible to reconstruct the analog signal from the IFC pulses with an error as small as required. In this paper, we present the definition of multiplication in pulse trains created by the IFC based on time domain operations and prove that it constitutes an Abelian group in the space of IFC pulse trains. We also show that pulse domain multiplication corresponds to pointwise multiplication of analog signals. It is further proved that pulse domain multiplication is distributive over pulse domain addition and hence it forms a field in the space of IFC pulse trains, which is an important property for linear signal processing.

## I. INTRODUCTION

This paper addresses multiplication of +/- unitary pulse trains created by the integrate and fire converter (IFC). A companion paper shows that it is also possible to define addition over pulse trains creating an Abelian group defined on pulse trains created by the IFC [1]. Together these papers show that it is possible to define a field over the set of pulse trains created by the IFC, which is rather important for signal processing. In fact, this opens up the possibility of operating algebraically with continuous time waveforms and creates an alternative for digital signal processing. The properties of the IFC are critical to define both operations since it creates an injective mapping between the two domains [2], [3]. In particular, the IFC approximates with an error as small as prescribed any band limited continuous time analog signal by rectangles of constant area [2]. Whenever the capacitor of the IFC reaches a threshold, a pulse is created and the capacitor is reset. Therefore the IFC approximates an "ideal" amplitude to time converter, and the time between pulses is proportional to the amplitude via the threshold [4]. Therefore, addition or multiplication of pulse trains becomes an addition of areas or multiplication of areas. The difficulty is that pulses must be created at the correct times to comply with the constant area constraint.

This paper follows basically the same procedure used in the companion addition paper to specify the operations required to multiply pulse trains, and to prove its properties. For completeness the concepts such as pulse train, constant area, and assumptions used in the companion paper that are common to both addition and multiplication are presented as an appendix.

The rest of the paper is organized as follows: Section II describes the theoretical foundations and proposes theorems for performing multiplication in the pulse domain. In section III, we study various group axioms and show that pulse domain multiplication constitutes an Abelian group and together with pulse domain addition forms a field in the space of IFC pulse trains. In section IV, we numerically evaluate the theorems and study the behavior of pulse domain multiplication under variation of different parameters.

## II. ALGEBRA ON IFC PULSE TRAINS: MULTIPLICATION

In IFC pulse trains, the area between consecutive pulse timings is equal to the threshold $\theta$; therefore operating algebraically with inter-pulse intervals (IPI) is equivalent to operating algebraically with the amplitude of the analog signal. In our companion paper [1], we showed that by relating the time between pulses to area, pulse trains can be added and the result corresponds to pointwise addition of the input analog signal.

The goal of this paper is to perform direct multiplication of pulse trains and show that it corresponds to pointwise multiplication of analog signals. The proposed pulse domain multiplication relies on the fact that the time between two pulses after multiplication contracts or expands based on the identity element i.e., if the time between pulses of the multiplier corresponds to a number less than one under the analog curve, then it leads to time expansion in the output and vice-versa. Therefore to multiply the two pulse trains, it is necessary to relate the IPI of the multiplier to a reference of one under the analog curve. Moreover, similar to pulse domain addition, it is imperative to quantify carryovers between subsequent evaluations [1]. Therefore, we start with preliminary definitions that will allow the implementation of the multiplication in pulse trains.

G. Nallathambi is with the Department of Electrical and Computer Engineering, Gainesville, Florida, 32611, USA (e-mail: gabriel_n,@ymail.com).

J. C. Principe is with the Department of Electrical and Computer Engineering, Gainesville, Florida, 32611, USA (e-mail: principe@cnel.ufl.edu).

## A. Definitions

*Reference pulse train:* It is a periodic pulse train i.e., the inter pulse intervals are equal and it corresponds to a constant value of 1 under the analog curve as shown in Fig. 1(a). It serves as the identity pulse train in multiplication.

*Relative multiplier area (RMA):* It is given by the ratio of the reference pulse interval to the multiplier pulse interval. Its value determines whether interval corresponds to a value greater than or less than 1 under the analog curve.

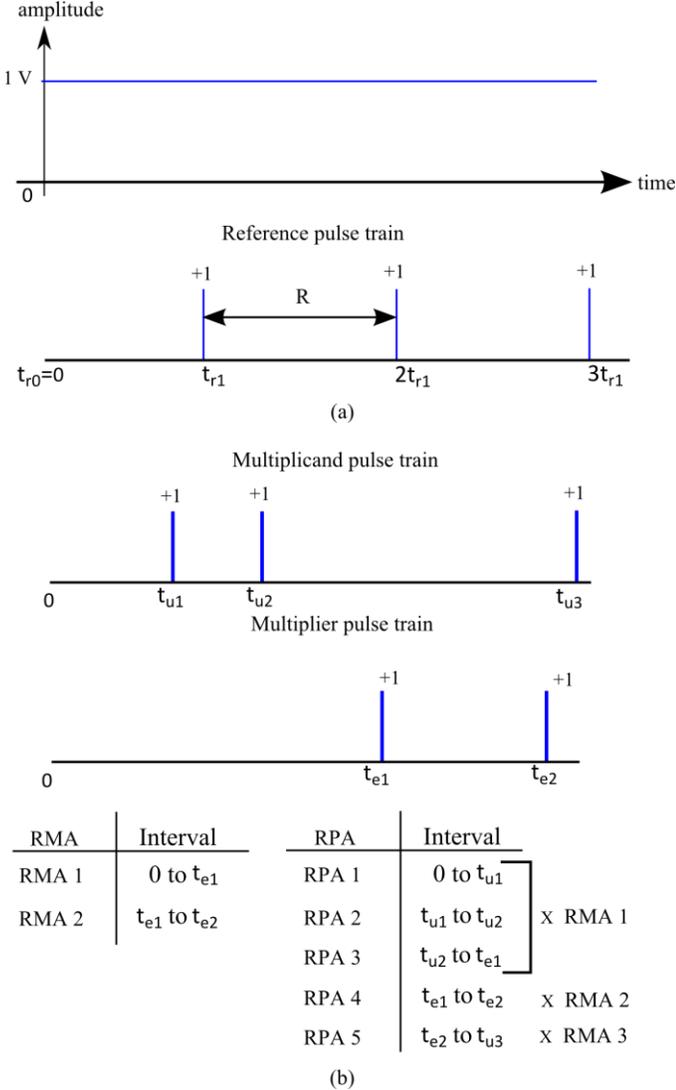

Fig. 1. (a) Periodic reference pulse train corresponds to constant 1 under the analog curve. (b) Interval wise pulse timing multiplication with respect to multiplier pulse train.

*Resultant product area (RPA):* RPA is the product of RMA and multiplicand areas in a given multiplier interval. It yields the total constant area in an interval due to multiplication of operands.

*Excess area:* It is the fraction of the constant area that remains after the occurrence of one or more pulses in the output due to multiplication. The excess area that remains is carried over to the next interval. It is given by: Net sum area − ⌊Net sum area⌋, where ⌊.⌋ is the floor operator.

*Net sum area (NSA):* NSA of an interval is given by the sum of excess area and RPA. It represents the total constant area after the occurrence of the last pulse in the output due to multiplication. The number of pulses in an interval resulting from multiplication is given by ⌊Net sum area⌋.

## B. Assumptions

*Assumptions 1 and 2:* The assumption of constant linear increase in area within every pulse interval and the refractory period being zero is common for both addition and multiplication. These assumptions were discussed in the companion paper [1] and are presented again in Appendix B.

*Assumption 3:* The computation of areas is done at every pulse timing with respect to the multiplier pulse train. This ensures the satisfaction of the constraint that the time between pulses represents a constant area. For instance, consider the multiplication of two pulse trains namely multiplicand, and multiplier pulse trains shown in Fig. 1(b). The first pulse of the multiplier pulse train occurs at $t_{e_1}$ and hence the first RMA is computed between 0 to $t_{e_1}$. But, in the same interval there are two pulses in the multiplicand; therefore, areas are computed in the intervals 0 to $t_{u_1}$, $t_{u_1}$ to $t_{u_2}$ and $t_{u_2}$ to $t_{e_1}$ and each of these areas are multiplied with the first RMA to give the respective RPA's and so on. Thus, interval wise multiplication is done with respect to the multiplier pulse train for calculating RMA and at every other pulse timing for calculating RPA.

## C. Theorem 1: Multiplication of pulse trains with amplitude >1.

Let $t_{r_1}$ be the first pulse of the periodic reference pulse train, i.e., periodicity, $R = t_{r_1}$. Let the *multiplicand* $P_1$ and the *multiplier* $P_2$ pulse trains have pulses with positive polarities at $t_{c_1}$ and $t_{e_1}, t_{e_2}, \cdots t_{e_m}$ respectively such that $t_{r_1} > t_{c_1} = t_{e_m} > \cdots > t_{e_2} > t_{e_1}$; then the resultant pulse train from the product of the multiplicand and multiplier pulse trains has pulses with positive polarities. Suppose their corresponding pulse timings are given by $T_k^{(j)}$, where $k$ is the $k^{th}$ pulse time and $n_j$ is the number of pulses in the interval $(t_{e_{j-1}}, t_{e_j})$, then

$$T_k^{(j)} = t_{e_{j-1}} + \frac{E_j}{R}\left[kC_1 - (j-1)R + C_1 \sum_{i=1}^{j-1} n_i\right] \quad \text{for} \quad j = 1,2,3,\cdots m \quad \text{and} \quad k = 1,2,3,\cdots n_j \quad \text{where} \quad n_j = \left\lfloor \frac{j.R - C_1 \sum_{i=1}^{j-1} n_i}{C_1} \right\rfloor, C_1 = t_{c_1}, E_i = (t_{e_i} - t_{e_{i-1}}), t_{e_0} = 0 \text{ and } n_0 = 0$$

with $C_1$ and $E_i$ being the inter-pulse multiplicand and multiplier duration respectively.

*Proof:*

Consider the multiplicand and multiplier pulse train shown in Fig. 2(a). In the interval $(0, t_{e_1})$, the rate of area per unit time in the multiplicand and multiplier are $\frac{1}{C_1}$ and $\frac{1}{E_1}$ respectively. The RMA is $\frac{R}{E_1}$ and hence the RPA and NSA in $(0, t_{e_1})$ is $\frac{R}{C_1}$, which is greater than one constant area. The # pulses in the first interval $(0, t_{e_1})$, $n_1$ is given by

$$n_1 = \left\lfloor \frac{R}{C_1} \right\rfloor \tag{1}$$

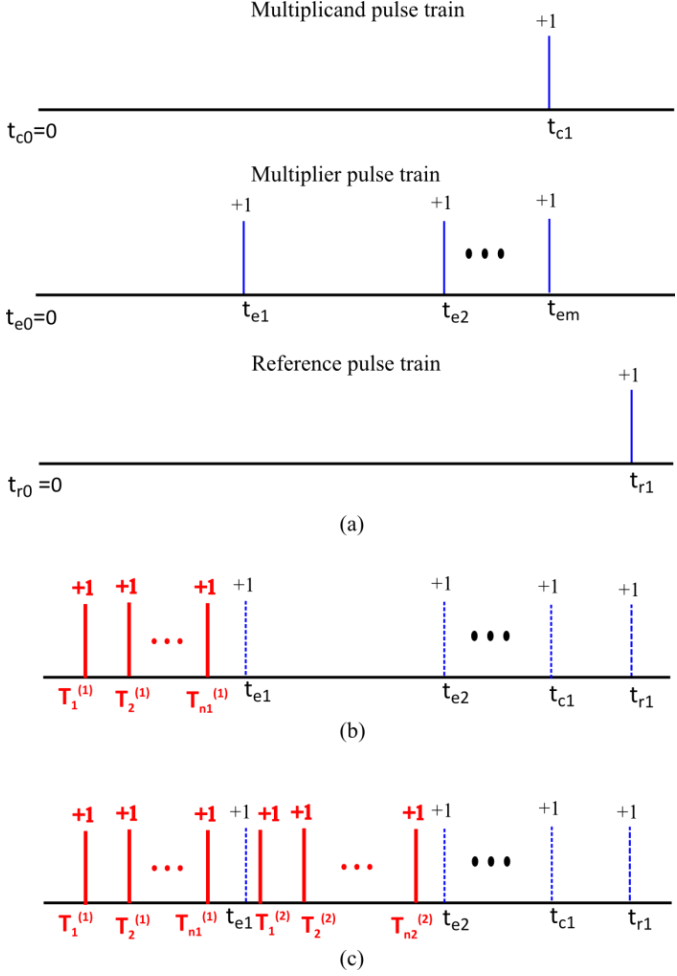

Fig. 2. Multiplication of pulse trains in theorem 1. (a) Multiplicand, multiplier and reference pulse trains. (b) Relative position of pulses in the first interval of resultant product with respect to other pulse timings. (c) Relative position of pulses in second interval with relation to other pulse timings

Let $T_1^{(1)}, T_2^{(1)} \cdots T_{n_1}^{(1)}$ denote the pulse timings in the first interval $(0, t_{e_1})$, then we have $T_1^{(1)} = \frac{time\ duration}{RPA\ in\ (0,t_{e_1})} = \frac{E_1 C_1}{R}$; $T_2^{(1)} = T_1^{(1)} + \frac{E_1 C_1}{R} = \frac{2E_1 C_1}{R}$; $\cdots T_{n_1}^{(1)}$ such that

$$T_{n_1}^{(1)} = n_1 \frac{E_1 C_1}{R} \quad (2)$$

the resultant pulse train at this stage is shown in Fig. 2(b) and we have

$$\text{Excess area in } (T_{n_1}^{(1)}, t_{e_1}) = \frac{R}{C_1} - n_1 \quad (3)$$

$$\text{Excess area time in } (T_{n_1}^{(1)}, t_{e_1}) = t_{e_1} - T_{n_1}^{(1)} \quad (4)$$

In the next interval $(t_{e_1}, t_{e_2})$, the rate of area per unit time in the multiplicand and multiplier are $\frac{1}{C_1}$ and $\frac{1}{E_2}$ respectively. The RMA is $\frac{R}{E_2}$ and the RPA in $(t_{e_1}, t_{e_2})$ is $\frac{R}{C_1}$, which is greater than one constant area. The NSA in $(T_{n_1}^{(1)}, t_{e_2})$ is given by the sum of excess area in $(T_{n_1}^{(1)}, t_{e_1})$ and the RPA in $(t_{e_1}, t_{e_2})$. From eqn. 3, we have

$$\text{NSA in } (T_{n_1}^{(1)}, t_{e_2}) = \frac{R}{C_1} + \left(\frac{R}{C_1} - n_1\right) = \frac{2R}{C_1} - n_1 \quad (5)$$

The # pulses in the second interval $(t_{e_1}, t_{e_2})$, $n_2$ is given by

$$n_2 = \left\lfloor \frac{2R}{C_1} - n_1 \right\rfloor \quad (6)$$

Let $T_1^{(2)}, T_2^{(2)} \cdots T_{n_2}^{(2)}$ denote the pulse timings in the second interval $(t_{e_1}, t_{e_2})$. The pulse time interval $T_1^{(2)} - T_{n_1}^{(1)}$ that represents one constant area is given by the sum of excess area time in $(T_{n_1}^{(1)}, t_{e_1})$ and the time for $(1 - excess\ area)$ in $(t_{e_1}, t_{e_2})$.

$$T_1^{(2)} = t_{e_1} + \frac{E_2(C_1 - R + C_1 n_1)}{R} \quad (7)$$

The next pulse time is given by $T_2^{(2)} = T_1^{(2)} + \frac{E_2 C_1}{R} = t_{e_1} + \frac{E_2(2C_1 - R + C_1 n_1)}{R}$. Generalizing in the interval $(t_{e_1}, t_{e_2})$, we have

$$T_{n_2}^{(2)} = t_{e_1} + \frac{E_2(n_2 C_1 - R + C_1 n_1)}{R} \quad (8)$$

The resultant pulse train at this stage is shown in Fig. 2(c) and we have

$$\text{Excess area in } (T_{n_2}^{(2)}, t_{e_2}) = \frac{2R}{C_1} - n_1 - n_2 \quad (9)$$

$$\text{Excess area time in } (T_{n_2}^{(2)}, t_{e_2}) = t_{e_1} - T_{n_1}^{(1)} \quad (10)$$

Repeating this procedure in the next interval $(t_{e_2}, t_{e_3})$, the RMA is $\frac{R}{E_3}$ and the RPA in $(t_{e_2}, t_{e_3})$ is $\frac{R}{C_1}$, which is greater than one constant area. The NSA in $(T_{n_2}^{(2)}, t_{e_3})$ is given by $\frac{3R - (n_1 + n_2)C_1}{C_1}$ and hence the # pulses in the interval $(t_{e_2}, t_{e_3})$, $n_3$ is $\left\lfloor \frac{3R - (n_1 + n_2)C_1}{C_1} \right\rfloor$. Let $T_1^{(3)}, T_2^{(3)} \cdots T_{n_3}^{(3)}$ denote the pulse timings in the interval $(t_{e_2}, t_{e_3})$, then we have $T_1^{(3)} = t_{e_2} + \frac{E_3[C_1 - 2R + C_1(n_1 + n_2)]}{R}$; $T_2^{(3)} = T_1^{(3)} + \frac{E_3 C_1}{R}$; $\cdots T_{n_3}^{(3)}$ such that

$$T_{n_3}^{(3)} = t_{e_2} + \frac{E_3[n_3 C_1 - 2R + C_1(n_1 + n_2)]}{R} \quad (11)$$

By induction, equations 2, 8, and 11 can be generalized as

$$T_k^{(j)} = t_{e_{j-1}} + \frac{E_j[kC_1 - (j-1)R + C_1 \sum_{i=1}^{j-1} n_i]}{R} \quad (12)$$

for $j = 1, 2, 3, \cdots m$ and $k = 1, 2, 3, \cdots n_j$ where $n_j = \left\lfloor \frac{j.R - C_1 \sum_{i=1}^{j-1} n_i}{C_1} \right\rfloor$, $C_1 = t_{c_1}$, $E_i = (t_{e_i} - t_{e_{i-1}})$, $t_{e_0} = 0$ and $n_0 = 0$.

*Corollary 1*

Let the multiplicand $P_1$, the multiplier $P_2$ and periodic reference pulse trains be as defined in theorem 1 such that $t_{r_1} > t_{c_1} = t_{e_m} > \cdots > t_{e_2} > t_{e_1}$; then the resultant pulse train from the product of the $P_1$ and $P_2$ is the same as the pulse train resulting from the product of $P_2$ and $P_1$ i.e., multiplication of pulse trains is commutative.

*Proof.*

Consider $P_1$ as multiplier and $P_2$ as multiplicand pulse trains. In the interval $(0, t_{c_1})$, the RMA is $\frac{R}{C_1}$. In the same interval, there are $m$ pulses in the multiplicand; therefore, each of the $m$ multiplicand pulse intervals has to be multiplied by the RMA. In the interval $(0, t_{e_1})$, the multiplicand represents one constant area and hence the RPA and NSA in $(0, t_{e_1}) = \frac{R}{C_1}$ is greater than one constant area. The # pulses in the first interval $(0, t_{e_1})$, $n_1$ is given by $n_1 = \left\lfloor \frac{R}{C_1} \right\rfloor$ and it is straightforward to show that $T_{n_1}^{(1)} = n_1 \frac{E_1 C_1}{R}$. From theorem 1 it is evident that the pulse timings for product of $P_1$ and $P_2$ is the same as the pulse timings resulting from the product of $P_2$ and $P_1$ in the interval $(0, t_{e_1})$.

In the next interval $(t_{e_1}, t_{e_2})$, the RMA is $\frac{R}{t_{c_1}}$, multiplicand represents one constant area and hence the RPA is $\frac{R}{C_1}$. The NSA in $(T_{n_1}^{(1)}, t_{e_2})$ is given by $\frac{2R}{C_1} - n_1$ which is greater than one constant area and it can be easily shown that $T_{n_2}^{(2)} = t_{e_1} + \frac{E_2(n_2 C_1 - R + C_1 n_1)}{R}$. Again, from theorem 1 we see that the pulse timings for the product of $P_1$ and $P_2$ is the same as the pulse timings resulting from the product of $P_2$ and $P_1$ in the interval $(t_{e_1}, t_{e_2})$.

Similarly, for the other intervals $(t_{e_{i-1}}, t_{e_i}), i = 3, 4, 5, \cdots m$ the pulse timings for $P_1.P_2$ are the same as that for $P_2.P_1$. Thus commutative property holds for multiplication of pulse trains when $t_{r_1} > t_{c_1} = t_{e_m} > \cdots > t_{e_2} > t_{e_1}$.

### D. Theorem 2: Multiplication of pulse trains of at least one operand with amplitude <1

Let $t_{r_1}$ be the first pulse of the periodic reference pulse train, i.e., periodicity, $R = t_{r_1}$. Let the multiplicand $P_1$ and the multiplier $P_2$ pulse trains have pulses with positive polarities at $t_{c_1}$ and $t_{e_1}, t_{e_2}, \cdots t_{e_m}$ respectively such that $t_{c_1} > t_{r_1} > t_{e_m} > \cdots > t_{e_2} > t_{e_1}$; then the resultant pulse train from the product of the multiplicand and multiplier pulse trains has no pulses in the interval $(0, t_{e_1})$ and in the interval $(t_{e_{j-1}}, t_{e_j})$ there may be zero or at most one pulse for $j = 2, 3, 4, \cdots m$.

*Proof.*

Consider the multiplicand and multiplier pulse train shown in Fig. 3(a). In the interval $(0, t_{e_1})$, the RPA and NSA in $(0, t_{e_1})$ is $\frac{R}{C_1}$, which is less than one constant area. Hence, the # pulses in the first interval $(0, t_{e_1})$ is zero and the excess area is given by $\frac{R}{C_1}$.

In the next interval $(t_{e_1}, t_{e_2})$, RPA in $(t_{e_1}, t_{e_2})$ is $\frac{R}{C_1}$ and we have

$$\text{NSA in } (t_{e_1}, t_{e_2}) = \frac{2R}{C_1} \tag{13}$$

The condition in eqn. 13 can have two cases as follows
*Case I (a) in* $(t_{e_1}, t_{e_2})$: $\frac{2R}{C_1} < 1$

In this case, there are no pulses in the interval $(t_{e_1}, t_{e_2})$ and we have

$$\text{Excess area in } (t_{e_1}, t_{e_2}) = \frac{2R}{C_1} \tag{14}$$

$$\text{Excess area time in } (t_{e_1}, t_{e_2}) = t_{e_2} \tag{15}$$

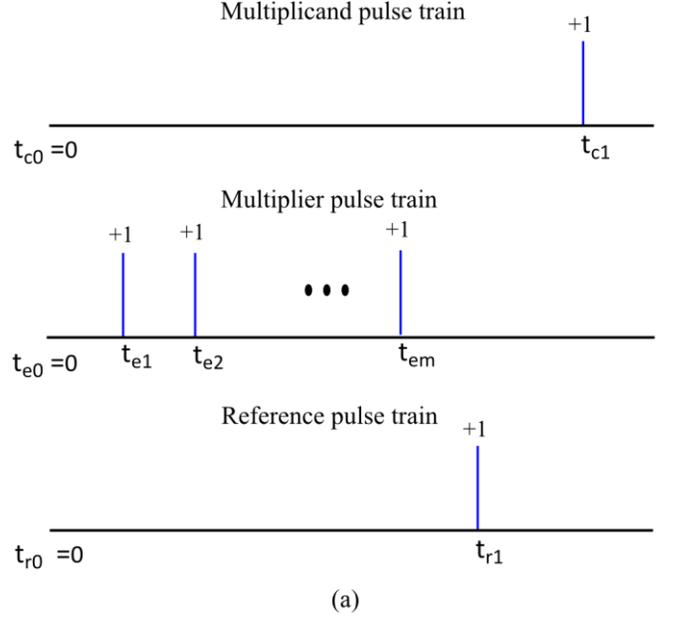

(a)

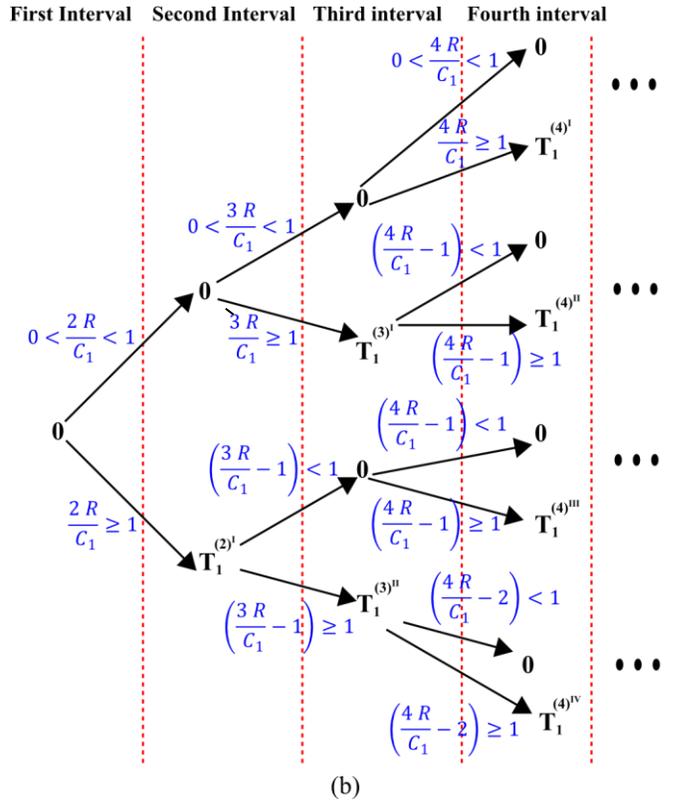

(b)

Fig. 3. Multiplication of pulse trains in theorem 2. (a) Multiplicand, multiplier and reference pulse trains. (b) Pulse sequence diagram.

*Case I (b) in* $(t_{e_1}, t_{e_2})$: $\frac{2R}{C_1} \geq 1$

In this case, one pulse occurs in the interval $(t_{e_1}, t_{e_2})$. Let $T_i^{(n)^J}$ denote the occurrence of $i$ pulses in the $n^{th}$ interval and $J$ represents the condition sequence of the pulse in the interval. The first pulse in the second interval $(t_{e_1}, t_{e_2})$, $T_1^{(2)I}$ is given by the sum of excess area time in $(0, t_{e_1})$ and $(1 - excess\ area)$ time in $(t_{e_1}, t_{e_2})$, and we have

$$T_1^{(2)I} = t_{e_1} + \frac{E_2}{R}(C_1 - R) \tag{16}$$

$$\text{Excess area in } (t_{e_1}, t_{e_2}) = \frac{2R}{C_1} - 1 \tag{17}$$

$$\text{Excess area time in } (t_{e_1}, t_{e_2}) = t_{e_2} - T_1^{(2)I} \tag{18}$$

Thus, there is zero or at most one pulse in the interval $(t_{e_1}, t_{e_2})$, and the sequence of pulses is shown in Fig. 3(b).

In the interval $(t_{e_2}, t_{e_3})$, RPA in interval $E_3$ is $\frac{R}{C_1}$ and the NSA in $E_3$ can take on two conditions resulting from equations 14 and 17 given by $\frac{3R}{C_1}$ and $\frac{3R}{C_1} - 1$ respectively. Condition I has two cases as follows:

Case I (a) in $(t_{e_2}, t_{e_3})$: $0 < \frac{3R}{C_1} < 1$

In this case, there are no pulses in the interval $(t_{e_2}, t_{e_3})$ and we have

$$\text{Excess area in } (t_{e_2}, t_{e_3}) = \frac{3R}{C_1} \tag{19}$$

$$\text{Excess area time in } (t_{e_2}, t_{e_3}) = t_{e_3} \tag{20}$$

Case I (b) in $(t_{e_2}, t_{e_3})$: $\frac{3R}{C_1} \geq 1$

In this case, one pulse occurs in the interval $(t_{e_2}, t_{e_3})$, and its pulse time $T_1^{(3)I}$ is given by

$$T_1^{(3)I} = t_{e_2} + \frac{E_3(C_1 - 2R)}{R} \tag{21}$$

$$\text{Excess area in } (t_{e_2}, t_{e_3}) = \frac{3R}{C_1} - 1 \tag{22}$$

$$\text{Excess area time in } (t_{e_2}, t_{e_3}) = t_{e_3} - T_1^{(3)I} \tag{23}$$

Similarly, from condition II, we have two cases as follows:

Case II (a) in $(t_{e_2}, t_{e_3})$: $\frac{3R}{C_1} - 1 < 1$

In this case, there are no pulses in the interval $(t_{e_2}, t_{e_3})$ and we have

$$\text{Excess area in } (t_{e_2}, t_{e_3}) = \frac{3R}{C_1} - 1 \tag{24}$$

$$\text{Excess area time in } (t_{e_2}, t_{e_3}) = t_{e_3} - T_1^{(2)I} \tag{25}$$

Case II (b) in $(t_{e_2}, t_{e_3})$: $\frac{3R}{C_1} - 1 \geq 1$

In this case, one pulse occurs in the interval $(t_{e_2}, t_{e_3})$, and its pulse time $T_1^{(3)II}$ is given by

$$T_1^{(3)II} = t_{e_2} - T_1^{(2)I} + \frac{E_3(2C_1 - 2R)}{R} \tag{26}$$

$$\text{Excess area in } (t_{e_2}, t_{e_3}) = \frac{3R}{C_1} - 2 \tag{27}$$

$$\text{Excess area time in } (t_{e_2}, t_{e_3}) = t_{e_3} - T_1^{(3)II} \tag{28}$$

Thus, there is zero or at most one pulse in the interval $(t_{e_2}, t_{e_3})$ and the sequence of pulses is shown in Fig. 3(b). Similarly, it is easy to show there is zero or one pulse in the interval $(t_{e_3}, t_{e_4})$ and the corresponding sequence transitions are shown in Fig. 3(b). Therefore, in the interval $(t_{e_{j-1}}, t_{e_j})$ there may be zero or at most one pulse for $j = 2, 3, 4, \cdots m$.

*Corollary 2*

Let the multiplicand $P_1$, the multiplier $P_2$ and periodic reference pulse trains be as defined in theorem 1 such that $t_{c_1} > t_{r_1} > t_{e_m} > \cdots > t_{e_2} > t_{e_1}$; then the resultant pulse train from the product of the $P_1$ and $P_2$ is the same as the pulse train resulting from the product of $P_2$ and $P_1$ i.e., multiplication of pulse trains is commutative.

*Corollary 3*

Let the multiplicand $P_1$ and the multiplier $P_2$ pulse trains have pulses with positive polarities at $t_{c_1}$ and $t_{e_1}, t_{e_2}, \cdots t_{e_m}$ respectively such that $t_{c_1} > t_{e_m} > \cdots > t_{e_2} > t_{e_1} > t_{r_1}$; then the resultant pulse train from the product of the multiplicand and multiplier pulse trains has no pulses in the interval $(0, t_{e_1})$ and in the interval $(t_{e_{j-1}}, t_{e_j})$ there may be zero or at most one pulse for $j = 2, 3, 4, \cdots m$.

*Remark:* To multiply pulse trains with positive and negative polarities, the aforementioned procedures are followed to obtain the pulse timings and the polarity of the pulses are given by bitwise multiplication of the individual polarities.

III. PROPERTIES OF MULTIPLICATION

We have shown that pulse train multiplication is commutative through corollary 1 and corollary 2. In this section, we will derive the other properties of pulse trains such as identity, inverse, associative and distributive laws.

*A. Theorem 3: Identity Element for multiplication*

Let $P_I$ denote the pulse train with pulses of positive polarity at $t_{r_1}, 2t_{r_1}, 3t_{r_1}, \cdots k t_{r_1}$ and $P$ be another pulse train with pulse at $t_{c_1}$. Suppose the periodic reference pulse train has periodicity $t_{r_1}$, then $P. P_I = P = P_I. P$.

*Proof.*

Case 1: $t_{r_1} > t c_1$

In the interval $(0, t_{c_1})$, the RMA and RPA is one constant area, and hence one pulse occurs at $t c_1$. Therefore the product of $P_I$ and $P$ is equal to $P$.

Case 2: $k\, t_{r_1} < t c_1$

In the interval $(0, t_{r_1})$, RPA and NSA due to the product of $P_I$ and $P$ is given by $\frac{R}{C_1}$ which is less than one constant area. Hence there is no pulse in $(0, t_{r_1})$ and the excess area is $\frac{R}{C_1}$. So $P_I. P = P$ in $(0, t_{r_1})$.

In the interval $(t_{r_1}, 2t_{r_1})$, NSA is given by $\frac{2R}{C_1}$ which is the same as eqn. 13. If $\frac{2R}{C_1} < 1$, no pulse occurs in $(t_{r_1}, 2t_{r_1})$; otherwise, one pulse occurs at $t_{c_1}$. Hence $P_I. P = P$ in $(t_{r_1}, 2t_{r_1})$.

Similarly, we can prove that $P_I . P = P$ in other pulse intervals. Since pulse multiplication is commutative, $P_I . P = P = P . P_I$.

B. *Theorem 4: Inverse element of pulse trains*

Let $P_1$ be a pulse train with pulses at $t_{c_1} = t_{r_1}, t_{c_2} = 2t_{r_1}, t_{c_3} = 3t_{r_1}, \cdots, t_{c_n} = nt_{r_1}$ and $P_2$ be another pulse train with pulses at $t_{e_1}, t_{e_2}, \cdots t_{e_m}$ such that $t_{r_1} > t_{e_m} > \cdots > t_{e_2} > t_{e_1}$. Suppose the periodic reference pulse train has periodicity $t_{r_1}$, then the pulse train $P_1/P_2$ is the inverse of $P_2$ and it is unique.

*Proof.*

Let $P_1$ be the multiplicand and $P_2$ be the multiplier pulse trains. Let $R = t_{r_1}$, $C_i = t_{c_i} - t_{c_{i-1}}$, $E_i = t_{e_i} - t_{e_{i-1}}$, $t_{e_0} = t_{c_0} = 0$. By the hypothesis, $R = C_1 = C_2 = C_3 = \cdots C_n$.

In the interval $(0, t_{e_1})$, the RMA of $P_2$ and the multiplicand area of $P_1$ is the same i.e., $\frac{R}{E_1}$. The RPA of the ratio $P_1/P_2$ is given by the product of multiplicand area and $1/RMA$. The RPA and NSA of $P_1/P_2$ in the interval $(0, t_{e_1})$ is $\frac{E_1^2}{R^2}$ which is less than one constant area. Hence there is no pulse in $(0, t_{e_1})$ and the excess area is $\frac{E_1^2}{R^2}$.

In the interval $(t_{e_1}, t_{e_2})$, the RPA of $P_1/P_2$ is $\frac{E_2^2}{R^2}$ and we have

$$\text{NSA in } (t_{e_1}, t_{e_2}) = \sum_{i=1}^{2} \frac{E_i^2}{R^2} \tag{29}$$

From the condition in eqn. 29, there are two cases as follows:

*Case I (a) in* $(t_{e_1}, t_{e_2})$: $\sum_{i=1}^{2} \frac{E_i^2}{R^2} < 1$

In this case, there are no pulses in the interval $(t_{e_1}, t_{e_2})$ and we have

$$\text{Excess area in } (t_{e_1}, t_{e_2}) = \sum_{i=1}^{2} \frac{E_i^2}{R^2} \tag{30}$$

$$\text{Excess area time in } (t_{e_1}, t_{e_2}) = t_{e_2} \tag{31}$$

*Case I (b) in* $(t_{e_1}, t_{e_2})$: $\sum_{i=1}^{2} \frac{E_i^2}{R^2} \geq 1$

In this case, one pulse occurs in the interval $(t_{e_1}, t_{e_2})$ and its timing $T_1^{(2)I}$ is given by

$$T_1^{(2)I} = t_{e_1} + \frac{R^2 - E_1^2}{E_2} \tag{32}$$

$$\text{Excess area in } (t_{e_1}, t_{e_2}) = \sum_{i=1}^{2} \frac{E_i^2}{R^2} - 1 \tag{33}$$

$$\text{Excess area time in } (t_{e_1}, t_{e_2}) = t_{e_2} - T_1^{(2)I} \tag{34}$$

In the interval $(t_{e_2}, t_{e_3})$, RPA in interval $E_3$ is $\frac{E_3^2}{R^2}$ and the NSA in $E_3$ can take on two conditions resulting from equations 30 and 33 given by $\sum_{i=1}^{3} \frac{E_i^2}{R^2}$ and $\sum_{i=1}^{3} \frac{E_i^2}{R^2} - 1$ respectively. Condition I results in two cases as follows:

*Case I (a) in* $(t_{e_2}, t_{e_3})$: $0 < \sum_{i=1}^{3} \frac{E_i^2}{R^2} < 1$

In this case, there are no pulses in the interval $(t_{e_2}, t_{e_3})$ and the excess area and excess area time in $(t_{e_2}, t_{e_3})$ are given by $\sum_{i=1}^{3} \frac{E_i^2}{R^2}$ and $t_{e_3}$ respectively.

*Case I (b) in* $(t_{e_2}, t_{e_3})$: $\sum_{i=1}^{3} \frac{E_i^2}{R^2} \geq 1$

In this case, one pulse occurs in the interval $(t_{e_2}, t_{e_3})$, and its pulse time $T_1^{(3)I}$ is given by

$$T_1^{(3)I} = t_{e_2} + \frac{R^2 - \sum_{i=1}^{2} \frac{E_i^2}{R^2}}{E_3} \tag{35}$$

The excess area and excess area time in $(t_{e_2}, t_{e_3})$ are given by $\sum_{i=1}^{3} \frac{E_i^2}{R^2} - 1$ and $t_{e_3} - T_1^{(3)I}$ respectively.

Condition II also results in two cases as follows:

*Case II (a) in* $(t_{e_2}, t_{e_3})$: $\sum_{i=1}^{3} \frac{E_i^2}{R^2} - 1 < 1$

In this case, there are no pulses in the interval $(t_{e_2}, t_{e_3})$ and the excess area and excess area time in $(t_{e_2}, t_{e_3})$ are given by $\sum_{i=1}^{3} \frac{E_i^2}{R^2} - 1$ and $t_{e_3} - T_1^{(2)I}$ respectively.

*Case II (b) in* $(t_{e_2}, t_{e_3})$: $\sum_{i=1}^{3} \frac{E_i^2}{R^2} - 1 \geq 1$

In this case, one pulse occurs in the interval $(t_{e_2}, t_{e_3})$, and its pulse time $T_1^{(3)II}$ is given by

$$T_1^{(3)II} = t_{e_2} - T_1^{(2)I} + \frac{2R^2 - \sum_{i=1}^{2} \frac{E_i^2}{R^2}}{E_3} \tag{36}$$

The excess area and excess area time in $(t_{e_2}, t_{e_3})$ are given by $\sum_{i=1}^{3} \frac{E_i^2}{R^2} - 2$ and $t_{e_3} - T_1^{(3)II}$ respectively. Similarly, the pulse timings in other intervals can be determined by following the same procedure. Suppose $P_3$ is another pulse train such that $P_3 . P_2 = P_I = P_2 . P_3$ where $P_I$ is the identity. Then, we have $P_3 = P_3 . P_I = P_3 . P_2 . (P_1/P_2) = P_I . (P_1/P_2) = (P_1/P_2)$. Thus $P_1/P_2$ is the inverse pulse train of $P_2$ and it is unique.

C. *Theorem 5: Associative property of multiplication for pulse trains*

Let $P_1$ be a pulse train with pulses at $t_{e_1}, t_{e_2}, \cdots t_{e_m}$, and the pulse trains $P_2$ and $P_3$ have pulses at $t_{c_1}$ and $t_{d_1}$ respectively. Suppose the periodic reference pulse train has periodicity $t_{r_1}$ such that $t_{r_1} > t_{d_1} > t_{c_1} = t_{e_m} > \cdots > t_{e_2} > t_{e_1}$, then $(P_1 . P_2) . P_3 = P_1 . (P_2 . P_3)$ in the interval $(t_{e_{j-1}}, t_{e_j})$ for $j = 1, 2, 3, \cdots m$ i.e., interval wise pulse train multiplication is associative.

*Proof.*

In the interval $(0, t_{e_1})$, the RMA of $P_1$ is $\frac{t_{r_1}}{t_{e_1}}$ and hence the RPA and NSA of $P_1 . P_2$ in $(0, t_{e_1})$ is $\frac{t_{r_1}}{t_{c_1}}$, which corresponds to a pulse timing given by $\frac{t_{e_1} t_{c_1}}{t_{r_1}}$. Now, the RMA of the pulse timing of $P_1 . P_2$ in $(0, t_{e_1})$ is given by $\frac{t_{r_1}^2}{t_{e_1} t_{c_1}}$ and we have

$$\text{NSA of } (P_1.P_2).P_3 \text{ in } (0, t_{e_1}) = \frac{t_{r_1}^2}{t_{c_1} t_{d_1}} \quad (37)$$

Likewise, the RPA and NSA of $P_2.P_3$ in $(0, t_{e_1})$ is given by $\frac{t_{r_1}}{t_{c_1}} \cdot \frac{t_{e_1}}{t_{d_1}}$ and we have

$$\text{NSA of } P_1.(P_2.P_3) \text{ in } (0, t_{e_1}) = \frac{t_{r_1}^2}{t_{c_1} t_{d_1}} \quad (38)$$

From equations 37 and 38, the NSA of $(P_1.P_2).P_3$ and $P_1.(P_2.P_3)$ are equal. Therefore in the interval $(0, t_{e_1})$ the pulse timings are equal for both $(P_1.P_2).P_3$ and $P_1.(P_2.P_3)$. Consequently, the excess area and excess area time are equal for $(P_1.P_2).P_3$ and $P_1.(P_2.P_3)$.

In the interval $(t_{e_1}, t_{e_2})$, the RMA of $P_1$ is $\frac{t_{r_1}}{t_{e_2}-t_{e_1}}$ and hence the RPA of $P_1.P_2$ in $(t_{e_1}, t_{e_2})$ is $\frac{t_{r_1}}{t_{c_1}}$, which corresponds to a pulse timing given by $\frac{(t_{e_2}-t_{e_1})t_{c_1}}{t_{r_1}}$. Now, the RMA of the pulse timing of $P_1.P_2$ in $(t_{e_1}, t_{e_2})$ is given by $\frac{t_{r_1}^2}{(t_{e_2}-t_{e_1})t_{c_1}}$ and we have

NSA of $(P_1.P_2).P_3$ in $(t_{e_1}, t_{e_2})$

$$= \text{Excess area in } (0, t_{e_1}) + \frac{t_{r_1}^2}{t_{c_1} t_{d_1}} \quad (39)$$

NSA of $P_1.(P_2.P_3)$ in $(t_{e_1}, t_{e_2})$

$$= \text{Excess area in } (0, t_{e_1}) + \frac{t_{r_1}^2}{t_{c_1} t_{d_1}} \quad (40)$$

From equations 39 and 40 it is evident that the NSA of $(P_1.P_2).P_3$ and $P_1.(P_2.P_3)$ are equal; therefore, in the interval $(t_{e_1}, t_{e_2})$ the pulse timings are equal for both $(P_1.P_2).P_3$ and $P_1.(P_2.P_3)$. This result can be proved for other pulse intervals. Thus interval wise pulse multiplication is associative.

### D. Theorem 6: Distributive property of multiplication over addition of pulse trains

Let $P_1$ be a pulse train with pulses at $t_{e_1}, t_{e_2}, \cdots t_{e_m}$, and the pulse trains $P_2$ and $P_3$ have pulses at $t_{c_1}$ and $t_{d_1}$ respectively. Suppose the periodic reference pulse train has periodicity $t_{r_1}$ such that $t_{r_1} > t_{d_1} > t_{c_1} = t_{e_m} > \cdots > t_{e_2} > t_{e_1}$, then in the interval $(t_{e_{j-1}}, t_{e_j})$, pulse multiplication is distributive over addition for $j = 1, 2, 3, \cdots m$ i.e.,

a. $(P_1 + P_2)P_3 = P_1P_3 + P_2P_3$ (Left distributive law)
b. $P_1(P_2 + P_3) = P_1P_2 + P_1P_3$ (Right distributive law)

*Proof.*

In the interval $(0, t_{e_1})$, the rate of area per unit time due to addition is given by the sum of rate of area per unit time in the

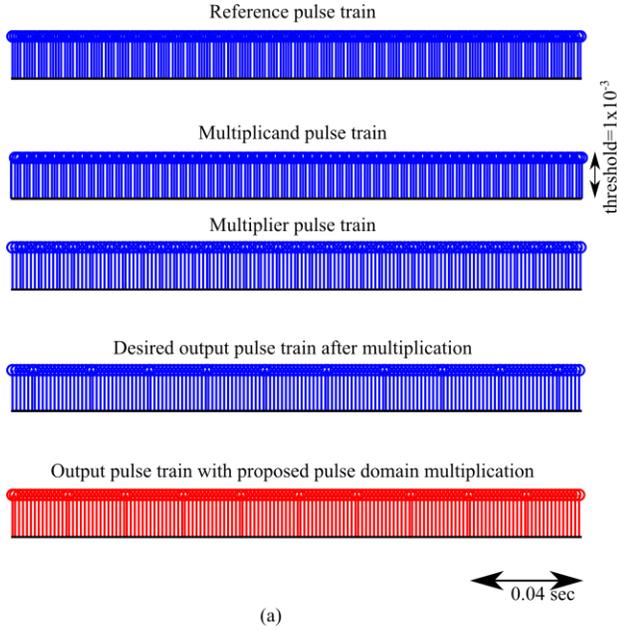
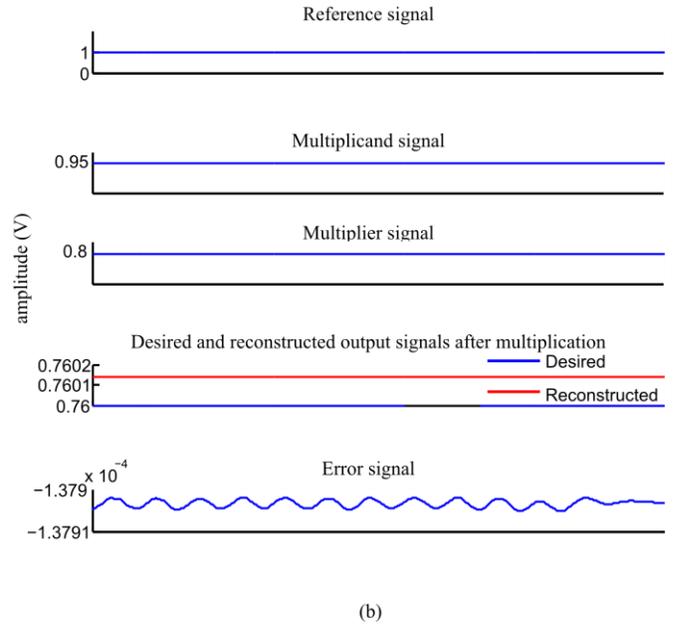

Fig. 4. Multiplication of periodic pulse trains. The left panel shows the multiplication of pulse trains corresponding to the signals in the right panel. (a) The reference, multiplicand, multiplier and desired output pulse trains are obtained with the threshold, leak factor, time stamping and refractory period set at 0.001, 40, 1µs and 0 respectively. The output pulse train calculated with the proposed pulse domain multiplication scheme is also shown. (b) The reference, multiplicand, multiplier and desired product of the signals that correspond to the pulse trains in the left panel are shown in the analog domain. The reconstructed signal of the output pulse train obtained with the proposed method and the error signal are shown in the bottom of the right panel. The SNR is 74.82dB.

Likewise, the RPA of $P_2.P_3$ in $(t_{e_1}, t_{e_2})$ is given by $\frac{t_{r_1}}{t_{c_1}} \cdot \frac{(t_{e_2}-t_{e_1})}{t_{d_1}}$ and we have

augend ($\frac{1}{t_{e_1}}$) and addend ($\frac{1}{t_{c_1}}$) respectively. The RMA of $P_1 +$

$P_2$ is $t_{r_1} \cdot \left( \frac{1}{t_{e_1}} + \frac{1}{t_{c_1}} \right)$, the multiplicand area of $P_3$ is $\frac{t_{e_1}}{t_{d_1}}$ and we have

$$\text{NSA of } (P_1 + P_2)P_3 \text{ in } (0, t_{e_1}) = t_{r_1} \cdot \left( \frac{t_{e_1} + t_{c_1}}{t_{c_1} t_{d_1}} \right) \quad (41)$$

Likewise, the RPA of $P_1 P_3$ and $P_2 P_3$ in $(0, t_{e_1})$ is given by $\frac{t_{r_1}}{t_{d_1}}$ and $\frac{t_{r_1} t_{e_1}}{t_{c_1} t_{d_1}}$ respectively and we have

$$\text{NSA of } P_1 P_3 + P_2 P_3 \text{ in } (0, t_{e_1}) = t_{r_1} \cdot \left( \frac{t_{e_1} + t_{c_1}}{t_{c_1} t_{d_1}} \right) \quad (42)$$

From equations 41 and 42, the NSA of $(P_1 + P_2)P_3$ and $P_1 P_3 + P_2 P_3$ are equal. Therefore in the interval $(0, t_{e_1})$ the pulse timings, the excess area and excess area time are equal for both $(P_1 + P_2)P_3$ and $P_1 P_3 + P_2 P_3$.

In the next interval $(t_{e_1}, t_{e_2})$, the RMA of $P_1 + P_2$ is $t_{r_1} \cdot \left( \frac{1}{t_{e_2} - t_{e_1}} + \frac{1}{t_{c_1}} \right)$, and we have

$$\begin{aligned}
\text{NSA of } (P_1 + P_2)P_3 &\text{ in } (t_{e_1}, t_{e_2}) \\
&= \text{Excess area in } (0, t_{e_1}) \\
&\quad + t_{r_1} \cdot \left( \frac{t_{e_2} - t_{e_1} + t_{c_1}}{t_{c_1} t_{d_1}} \right)
\end{aligned} \quad (43)$$

Moreover, the RPA of $P_1 P_3$ and $P_2 P_3$ in $(t_{e_1}, t_{e_2})$ is given by $\frac{t_{r_1}}{t_{d_1}}$ and $\frac{t_{r_1}(t_{e_2} - t_{e_1})}{t_{c_1} t_{d_1}}$ respectively and we have

$$\begin{aligned}
\text{NSA of } P_1 P_3 + P_2 P_3 &\text{ in } (t_{e_1}, t_{e_2}) \\
&= \text{Excess area in } (0, t_{e_1}) \\
&\quad + t_{r_1} \cdot \left( \frac{t_{e_2} - t_{e_1} + t_{c_1}}{t_{c_1} t_{d_1}} \right)
\end{aligned} \quad (44)$$

From equations 43 and 44, it is clear that # pulses, pulse timings, excess area and excess area time are equal for $(P_1 + P_2)P_3$ and $(P_1 P_3 + P_2 P_3)$ in $(t_{e_1}, t_{e_2})$.

Following the same procedure, it can be proved that $(P_1 + P_2)P_3 = (P_1 P_3 + P_2 P_3)$ in other pulse intervals. Hence left distributive law holds. Similarly, right distributive law can also be proved.

*Remark:* The set of all pulse trains forms a field under the binary operations namely interval wise pulse addition and pulse multiplication.

## IV. NUMERICAL RESULTS

To handle continuous time asynchronous pulses, a time stamping clock is used to quantize pulse times [1]. Similar to pulse domain addition, implementation issues results in timing imprecision that degrades the quality of multiplication of pulse trains just like finite wordlength degrades binary arithmetic. The signal-to-noise ratio (SNR) was used as the performance metric and is given by $10 \log \frac{P_{ds}}{P_{ds} - P_{rs}}$, where $P_{ds}$ is the power of the desired signal obtained by the product of the operands in the analog domain and $P_{rs}$ is the power of the reconstructed signal obtained by recovering the analog signal from the output pulse train. The effect of IFC threshold and clock variations on the SNR are also studied in line with the companion paper.

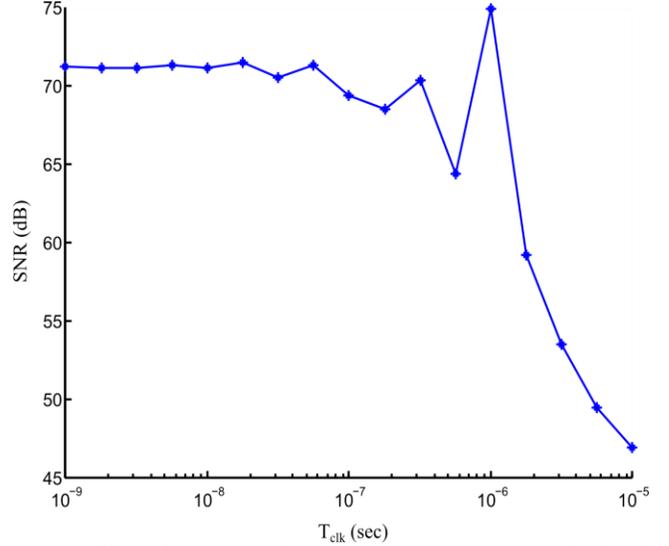
Fig. 5. Effect of time stamping on SNR. The SNR was calculated for the multiplicand and multiplier shown Fig. 4 with threshold, leak factor and refractory period set at 0.001, 40 and 0 respectively.

In Fig. 4, we study the multiplication of two periodic pulse trains that correspond to multiplicand=0.95V and multiplier=0.8V. Since both operands are less than 1V, the IPI of the desired output pulse train is greater than both operands. Like addition, the performance of the pulse domain multiplication scheme is limited by the timing imprecision in the output pulse train and the recovered signal is not exactly 0.76V in the analog domain resulting in an SNR of 74.82dB obtained with the threshold, leak factor, time stamping clock and refractory period set at 0.001, 40, 1μs and 0 respectively.

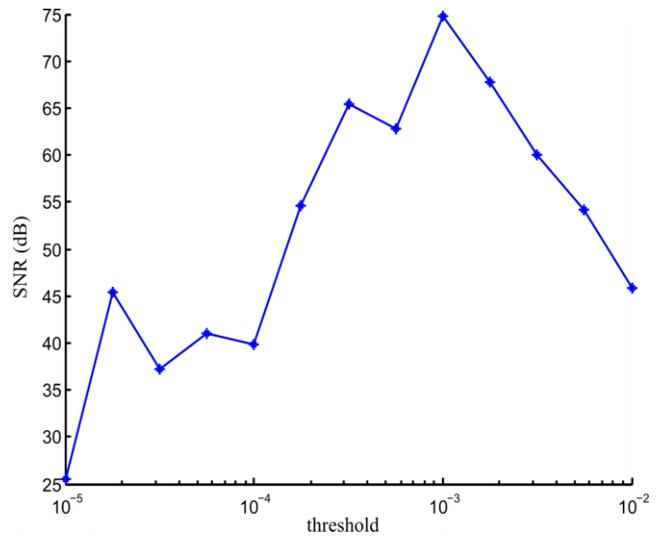
Fig. 6. Effect of threshold on SNR. The SNR was calculated for the multiplicand and multiplier shown in Fig. 4 with time stamping, leak factor and refractory period set at 1μs, 40 and 0 respectively

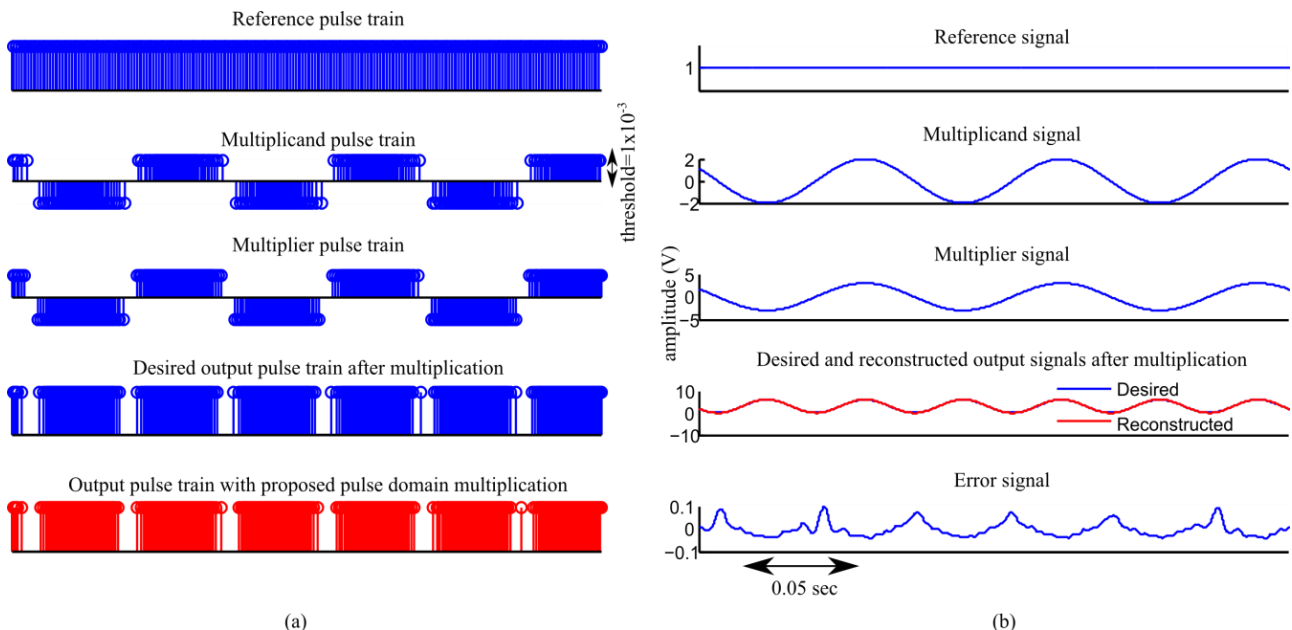

Fig. 7. Multiplication of aperiodic pulse trains. The left panel shows the multiplication of pulse trains corresponding to the signals in the right panel. (a) The reference, multiplicand, multiplier and desired output pulse trains are obtained with the threshold, leak factor, time stamping clock and refractory period set at 0.001, 40, 1μs and 0 respectively. The output pulse train calculated with the proposed pulse domain multiplication scheme is also shown. (b) The reference, multiplicand, multiplier and desired product of the 12 Hz sinusoidal signals that correspond to the pulse trains in the left panel are shown in the analog domain. The reconstructed signal of the output pulse train obtained with the proposed method and the error signal are shown in the bottom of the right panel. The SNR is 41.12 dB.

The behavior of the SNR with the time stamping clock for a simulated IFC pulse domain multiplication scheme is shown in Fig. 5. The input signal, threshold, leak factor and refractory period were chosen as in Fig. 4 and were kept constant to ensure a constant pulse rate. The variation of SNR with the clock is as expected, similar to pulse domain addition [1]. At higher clock frequencies, the SNR plateaus due to limited precision of the digital arithmetic while it degrades substantially at lower clock frequencies due to quantization of the pulse timings.

Then to study the variation of SNR with pulse rate the input signal, clock period, leak factor and refractory period were chosen as in Fig. 4. From Fig. 6, we see that at lower and higher thresholds, the SNR degrades due to limited precision of the digital arithmetic and quantization respectively. This behavior is consistent with pulse domain addition [1], resulting in similar trade-offs between error, pulse rate and choice of time stamping clocks for a given set of operands.

We also studied the multiplication of two aperiodic pulse trains that correspond to sinusoidal multiplicand and multiplier signals of $2\sin(24\pi t)$ and $3\sin(24\pi t)$ with the threshold, leak factor, clock period and refractory period set at 0.001, 40, 1μs and 0 respectively. With the proposed pulse domain multiplication method, the resulting SNR is 41.12dB. From Fig. 7, we see that the error is relatively high at the zero crossings when compared with other regions of the signal, which is analogous with pulse domain addition [1]. In our companion paper, we had shown that it is a direct consequence of assumption 1 and discussed the trade-offs associated with it, which will be studied theoretically in a follow-up paper. Also, these results can be easily extended to non-zero refractory period by following the same procedure described in the companion paper [1].

## V. Conclusion

In this paper, we present a methodology to perform multiplication of pulse trains and prove that it obeys the properties of an Abelian group. Pulse domain multiplication also forms a field over pulse domain addition, which is very important for linear signal processing. Analogous to pulse domain addition, all the operations are defined based on IPI; therefore it can be implemented online with a small delay given by the maximum IPI. We have also shown that, multiplication of pulse trains corresponds to pointwise multiplication of the corresponding analog signals that are fed to the pair of IFC. Hence we effectively propose a methodology to algebraically process continuous time, continuous amplitude signal sources.

Operating algebraically with continuous amplitude, continuous time signals is novel. In signal processing theory, the most fundamental operation is the inner product. In our future work, we would implement inner product for pulse trains using pulse domain addition and multiplication. Filtering using pulse trains will be a novelty for the signal processing community, and opens the door to processing continuous time signals based on algebra instead of physical electrical components.

APPENDIX A: DEFINITIONS

**Pulse train**: Let us define a pulse train P as an ordered sequence of events over time with positive (+) or negative (-) polarity such that $P = \{p_{t_k}\}$ where $t_k \in \Re, p = \pm 1, k = -\infty, \ldots, 0, 1, \ldots \infty$. Apart from polarity, which can be handled separately, the time between pulses is our main interest. Notice that P is an ordered set, so we cannot change the element order. In order to enforce the time evolution, let us further denote as $D_k$ the time between two consecutive events (left to right order), i.e. $D_k = p_{t_k} - p_{t_{k-1}}, D_k \in \Re^+, k = 1, 2, \ldots \infty$. Now $D_k$ are positive real numbers, and they constitute the underlying set for our field along the polarity of the corresponding pulses, i.e. the domain is $\Re$, the real numbers.

**Constant area:** Every pulse interval, $D_k$ resulting from the binary operations in pulse train must satisfy the constant area condition i.e., the area under the timing between the pulses should always be equal to one constant area. The number of constant areas resulting from binary operations governs the number of pulses in the output.

APPENDIX B: ASSUMPTIONS

**Assumption 1:** We assume that the rate of area per unit time within every pulse interval $D_k$ is constant i.e., the threshold $\theta$ (constant area) is reached linearly. This approximation is accurate for periodic pulse trains. In case of aperiodic pulse trains, the approximation is nearly accurate for small $D_k$, while if $D_k$, is large it results in an error that is a fraction of $\theta$. This assumption is necessary for finding the rate of area of the operands in a given time within an interval.

**Assumption 2:** For mathematical convenience, the refractory period is assumed to be zero. This does not affect the continuity of the proofs. Infact, the refractory period can be simply added to the resulting pulse timing without any error.